\documentclass[%
aip,
amsmath,amssymb,
reprint
]{revtex4-1}
\usepackage{graphicx}
\usepackage{dcolumn}
\usepackage{bm}

\usepackage{amsmath}
\usepackage{color}
\usepackage[normalem]{ulem}
\usepackage{balance}
\usepackage[T1]{fontenc}

\begin{document}

\preprint{AIP/123-QED}

\title{Unified Treatment for Scattering, Absorption, and Photoluminescence of coupled Metallic Nanoparticles with Vertical Polarized Excitation}

\author{Yuqing Cheng}\affiliation{1 School of Mathematics and Physics, Beijing Advanced Innovation Center for Materials Genome Engineering, University of Science and Technology Beijing, Beijing 100083, China}%
\author{Mengtao Sun}\thanks{mengtaosun@ustb.edu.cn}
\affiliation{1 School of Mathematics and Physics, Beijing Advanced Innovation Center for Materials Genome Engineering, University of Science and Technology Beijing, Beijing 100083, China}%


\begin{abstract}
Optical properties of coupled metallic particles (MNPs) have been widely reported due to their unique characteristics such as peak shift/splitting of the coupling spectra and electromagnetic enhancement at sub-wavelength scale, etc. In a previous work, we have investigated the coupling spectra of two coupled MNPs with parallel polarized excitation. In this study, we investigate the vertical polarization case in detail. Different from the parallel one, the vertical one has its unique properties: (a) three coupling coefficients; (b) positive coupling terms in the coupling equations; (c) blue-shifts of the peaks with the increasing coupling strength for identical MNPs spectra, including scattering, absorption, and photoluminescence. Comparison with published experimental results shows the validity of this model. This work provides a deeper understanding on the optical properties of coupled MNPs and is beneficial to relevant applications.
\end{abstract}

\maketitle


\section{\label{sec:Introduction}Introduction}
The optical properties of coupled metallic nanostructures, or metallic nanoparticles (MNPs) have attracted wide interest of researchers due to the ability and potential to control the electromagnetic field at subwavelength scale. Numerous applications have been driven by these studies, such as optical biosensing,\cite{appbio1, appbio2, appbio3, appbio4} chiral optics, \cite{appco1, appco2, appco3}, plasmonic devices, \cite{apppd1, apppd2, apppd3}, and nonlinear optics.\cite{appnlo1, appnlo2, appnlo3}

There are numerous coupling models using harmonic oscillator system.
Joe et al.\cite{Fano} used both the classical and quantum systems to describe the Fano resonance. In their classical model, two oscillators couple with each other with the external force driving one of them. The phase of these two are detailed analyzed to explain the formation of Fano line shape. In their quantum model, they analyzed the coupling between 2D electron waveguide with a quantum dot, and the Fano resonance results from the interference between a propagating wave and an evanescent wave.
Quantum oscillator models could explain the coupling in a similar way. Fan et al.\cite{modelFan} present a model to describe a single optical resonance coupled with multiple input and output ports, employing. Yang et al.\cite{modelYang} used the quantum oscillator model to explain electromagnetically induced transparency in their all-dielectric metasurface.
These examples that employ oscillator models are successful. However, some details are missing. Especially, the coupling coefficients are not clarified. That is, when considering a practical situation, e.g., MNPs with certain separate distance, the coupling coefficients are usually uncertain. Researchers often fit the curves of experiments to determine the coefficients in general. Furthermore, the coupling coefficients may vary with several quantities, thus influencing the coupling process. 

In a previous work, we have presented a practical model to investigate the coupled MNPs with parallel polarized excitation.
In this study, we also present a practical model to investigate the optical properties of two coupled MNPs with vertical polarized excitation. The results in this work show great difference from the parallel polarized one in at least three points: (a) There are three coupling coefficients for vertical polarization, while there are only two for parallel polarization; (b) vertical polarization has positive coupling terms in the coupling equations, while parallel polarization has negative ones; (c) for identical MNPs, as the coupling strength increases, vertical polarization shows blue-shift in the spectra, while parallel polarization shows red-shift. These phenomena are explained by this model.
This work would help to understand the coupling phenomena of MNPs in a classical way more deeply.

\section{\label{sec:Model}Model}
\begin{figure}[tb]
\includegraphics[width=0.48\textwidth]{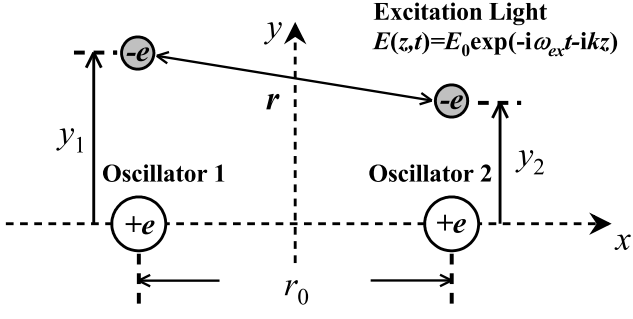}
\caption{\label{fig:Schematic} Schematic of the coupled MNPs with vertical polarized excitation. The two MNPs are on the $x$-axis; circles filled with gray stand for electrons with negative charge with distance $r$, and circles filled with white stand for ions with positive charge with distance $r_0$; $y_1$ and $y_2$ are the displacements from the equilibrium positions in $y$-direction of the two electrons, respectively; the excitation electric field is $y$-polarized with angular frequency $\omega_{ex}$.
}
\end{figure}

Fig. \ref{fig:Schematic} shows the schematic of this model in $x$-$y$ view. Two MNPs are treated as two oscillators with ions (positive charged) staying at rest and electrons (negative charged) oscillating around the ion. Here, the two oscillators are on the $x$-axis, and the distance between them is $r_0$. The $y$-polarized external electromagnetic field with angular frequency $\omega_{ex}$ and amplitude $E_0$ propagates along the $z$-axis and illustrates the coupled MNPs. Therefore, the polarization of the excitation light is perpendicular to the line between the two oscillators (vertical polarized excitation). We assume that the free electrons are forced to oscillate only along $y$-axis.
Define $y_j(t)$, $\dot{y}_j(t)$, and $\ddot{y}_j(t)$ as the displacement, velocity, and accelerate in $y$-direction of the $j$th electron with $j=1,2$, respectively. The equations of the two electrons are written as:
\begin{subequations}
\begin{align}
\ddot{y}_1+ \beta_{01} \dot{y}_1+\omega_{01}^2 y_1-\frac{F_{21}}{m_e}=C_1 \mathrm{exp}(-\mathrm{i} \omega_{ex} t), \label{eq:basic01a} \\
\ddot{y}_2+ \beta_{02} \dot{y}_2+\omega_{02}^2 y_2-\frac{F_{12}}{m_e}=C_2 \mathrm{exp}(-\mathrm{i} \omega_{ex} t). \label{eq:basic01b}
\end{align}
\label{eq:basic01}
\end{subequations}
Here, $m_e$ is the electron mass, $C_1=-e E_1/m_e$, $C_2=-e E_2/m_e$, $e$ is the elementary charge. Usually, $E_1=E_2=E_0$ is a good approximation due to the sub-wavelength distance between them; $\omega_{0j}$ and $\beta_{0j}$ are the eigenfrequency (angular) and the damping coefficient of the $j$th oscillator, respectively; $F_{21}=-e E_{21}$ and $F_{12}=-e E_{12}$ are the interaction forces between the two oscillators, which can be derived from:\cite{Griffiths}
\begin{equation}
\mathbf{E}=\frac{q}{4 \pi \varepsilon_0} \frac{r}{(\mathbf{r} \cdot \mathbf{u})^3} \left [ (c^2-v^2)\mathbf{u} + \mathbf{r} \times (\mathbf{u} \times \mathbf{a}) \right ],
\label{eq:E}
\end{equation}
where $\mathbf{E}$ is the electric field introduced by the moving charge $q$, $\varepsilon_0$ is the permittivity of vacuum, and $c$ is the velocity of light in vacuum; $\mathbf{u} \equiv c \mathbf{r} /r - \mathbf{v}$; $\mathbf{r}$, $\mathbf{v}$, and $\mathbf{a}$ are the displacement vector from the charge to the field point, velocity, and the accelerate of the charge, respectively, with $r=|\mathbf{r}|$ and $v=|\mathbf{v}|$. Therefore, the electric field $E_{21}$ and $E_{12}$ can be written as:
\begin{subequations}
\begin{align}
E_{21}\cong \frac{e}{4 \pi \varepsilon_0 r_0^2 }(\frac{y_2}{r_0}+\frac{\dot{y}_2}{c}+\frac{r_0\ddot{y}_2}{c^2}),
\label{eq:E2a} \\
E_{12}\cong \frac{e}{4 \pi \varepsilon_0 r_0^2 }(\frac{y_1}{r_0}+\frac{\dot{y}_1}{c}+\frac{r_0\ddot{y}_1}{c^2}).
\label{eq:E2b}
\end{align}
\label{eq:E2}
\end{subequations}
Here, we use the conditions $v/c \ll 1$, $y/r_0 \ll 1$, and $r_0\ddot{y}/c^2 \ll 1$, and ignore the higher-order infinitesimal for approximation. Due to the collective oscillation of large number of electrons in the MNPs, the interaction forces should be modified as $F_{21}=-N_2 e E_{21}$ and $F_{12}=-N_1 e E_{12}$, where $N_1$ and $N_2$ are the effective numbers of free electrons in Oscillator 1 and Oscillator 2, respectively. To make it clearer, we define the general coupling coefficients as:
\begin{equation}
\begin{aligned}
&\frac{1}{\kappa}=\frac{e^2}{4 \pi \varepsilon_0 m_e c^3},\\
&g_0^2=\frac{1}{\kappa}\left(\frac{c}{r_0}\right)^3,~
\gamma_0=\frac{1}{\kappa}\left(\frac{c}{r_0}\right)^2,~
\eta_0=\frac{1}{\kappa}\left(\frac{c}{r_0}\right)^1,
\label{eq:general}
\end{aligned}
\end{equation}
then the coupling coefficients are:
\begin{equation}
\begin{aligned}
g_{1}^2=N_1 g_0^2,~ \gamma_{1}=N_1 \gamma_0,~ \eta_{1}=N_1 \eta_0,\\
g_{2}^2=N_2 g_0^2,~ \gamma_{2}=N_2 \gamma_0,~ \eta_{2}=N_2 \eta_0,
\end{aligned}
\label{eq:g}
\end{equation}
and Eq. (\ref{eq:basic01}) can be written as:
\begin{subequations}
\begin{align}
\ddot{y}_1+ \beta_{01} \dot{y}_1+\omega_{01}^2 y_1+\eta_{2}\ddot{y}_2+\gamma_{2} \dot{y}_2+g_{2}^2 y_2=C_1 \mathrm{exp}(-\mathrm{i} \omega_{ex} t), \label{eq:basic1a} \\
\ddot{y}_2+ \beta_{02} \dot{y}_2+\omega_{02}^2 y_2+\eta_{1}\ddot{y}_1+\gamma_{1} \dot{y}_1+g_{1}^2 y_1=C_2 \mathrm{exp}(-\mathrm{i} \omega_{ex} t). \label{eq:basic1b}
\end{align}
\label{eq:basic1}
\end{subequations}
For simplicity, we define $\Omega_j(\alpha)=\omega_{0j}^2+\beta_{0j}\alpha+\alpha^2$ and $G_j(\alpha)=g_{j}^2+\gamma_j \alpha+\eta_{j} \alpha^2$ for $j=1,2$ which would be used in the following derivation.

\subsection{\label{sec:SCA}Scattering and Absorption}
To obtain the white light scattering spectra, $\alpha$ is substituted by $-\mathrm{i}\omega_{ex}$, and we solve the following equations derived from Eq. (\ref{eq:basic1}) after assuming $y_j(t)=A_j \mathrm{exp}(-\mathrm{i}\omega_{ex}t)$ for $j=1,2$,
\begin{equation}
\begin{aligned}
&
\begin{pmatrix}
\Omega_1(\alpha) & G_2(\alpha) \\
G_1(\alpha) & \Omega_2(\alpha)
\end{pmatrix}
\left(
\begin{array}{c}
A_1 \\
A_2 \\
\end{array}\right)
=\left(
\begin{array}{c}
C_1 \\
C_2 \\
\end{array}\right),
\\ & \mathrm{with}~ \alpha=-\mathrm{i}\omega_{ex}.
\end{aligned}
\label{eq:SCAm}
\end{equation}
The solutions are:
\begin{equation}
\begin{aligned}
A_1(\omega_{ex})=\frac{\Omega_2 C_1-G_2 C_2}{\Omega_1 \Omega_2 - G_1 G_2},\\
A_2(\omega_{ex})=\frac{\Omega_1 C_2-G_1 C_1}{\Omega_1 \Omega_2 - G_1 G_2}.
\end{aligned}
\label{eq:A}
\end{equation}
Therefore the total scattering spectrum is derived after substituting $\omega_{ex}$ with $\omega$:
\begin{equation}
\begin{aligned}
I_{sca}(\omega)=\omega^4 \left|N_1 A_1(\omega)+N_2 A_2(\omega) \right|^2.
\end{aligned}
\label{eq:Isca1}
\end{equation}
Here, the term $\omega^4$ is due to the fact that the detected scattering field is usually the far field, which is proportional to $\ddot{y} \propto \omega^2$.

To obtain the absorption spectrum,\cite{Fano} we notice that the absorption is introduced by the term $\beta_{0j}\dot{y}$, thus the total absorption spectrum:
\begin{equation}
\begin{aligned}
I_{abs}(\omega)=N_1\left| \beta_{01} A_1(\omega)\right|^2+N_2\left| \beta_{02} A_2(\omega) \right|^2.
\end{aligned}
\label{eq:Iabs}
\end{equation}

\subsection{\label{sec:PL}Photoluminescence}
To obtain the PL spectrum, we should find the find the eigen solutions of the following equations:
\begin{subequations}
\begin{align}
\ddot{y}_1+ \beta_{01} \dot{y}_1+\omega_{01}^2 y_1+\eta_{2}\ddot{y}_2+\gamma_{2} \dot{y}_2+g_{2}^2 y_2=0,
\label{eq:basic2a} \\
\ddot{y}_2+ \beta_{02} \dot{y}_2+\omega_{02}^2 y_2+\eta_{1}\ddot{y}_1+\gamma_{1} \dot{y}_1+g_{1}^2 y_1=0.
\label{eq:basic2b}
\end{align}
\label{eq:basic2}
\end{subequations}
After assuming $y_j(t)=B_j\mathrm{exp}(\alpha t)$ for $j=1,2$, the equations are written in matrix form:
\begin{equation}
\begin{aligned}
&
\begin{pmatrix}
\Omega_1(\alpha) & G_2(\alpha) \\
G_1(\alpha) & \Omega_2(\alpha)
\end{pmatrix}
\left(
\begin{array}{c}
B_1 \\
B_2 \\
\end{array}\right)
=\left(
\begin{array}{c}
0 \\
0 \\
\end{array}\right),
\\ & \mathrm{with~initial~conditions:}\\
& y_j(0)=A_j,~\dot{y}_j(0)=0, ~ \mathrm{for}~j=1,2
\end{aligned}
\label{eq:PLm}
\end{equation}
Obviously, to find a non-trivial solution, $\alpha$ should satisfy
\begin{equation}
\begin{vmatrix}
\Omega_1(\alpha) & G_2(\alpha) \\
G_1(\alpha) & \Omega_2(\alpha)
\end{vmatrix}
=0.
\label{eq:PLm2}
\end{equation}
A particular case is that the two oscillators are identical, i.e., $\omega_{01}=\omega_{02}=\omega_0$, $\beta_{01}=\beta_{02}=\beta_0$, $N_1=N_2=N$, $g_{1}=g_{2}=g$, $\gamma_1=\gamma_2=\gamma$, $\eta_1=\eta_2=\eta$, and $C_1=C_2=C_0$. In the rest of this section, we take this identical case as an example to illustrate the PL properties of the coupled system. The expressions derived from the general case (non-identical) are much more complicated than the identical one, but the solving processes of them are similar. Back to the identical case, the solutions of $\alpha$ are:
\begin{equation}
\begin{aligned}
\alpha_{p\pm}=\frac{-\beta_0-\gamma\pm\mathrm{i}\sqrt{4(1+\eta)(\omega_0^2+g^2)-(\beta_0+\gamma)^2}}{2(1+\eta)},\\
\alpha_{m\pm}=\frac{-\beta_0+\gamma\pm\mathrm{i}\sqrt{4(1-\eta)(\omega_0^2-g^2)-(\beta_0-\gamma)^2}}{2(1-\eta)}.
\end{aligned}
\label{eq:alpha1}
\end{equation}
Notice that these solutions should satisfy $\eta \ne 1$ (Case 1). If $\eta=1$ and $\gamma \ne \beta_0$ (Case 2), the solutions would be reduced to $\alpha_{p\pm}$ and $\alpha_m$, with $\alpha_m=\frac{\omega_0^2-g^2}{\gamma-\beta_0}$. In this case, $\alpha_m$ corresponds to the non-oscillation term with exponentially decreasing or increasing, depending on the sign of $\alpha_m$, the latter of which should be removed from the solutions due to its divergence with time. If $\eta=1$ and $\gamma=\beta_0$ (Case 3), the solutions would only remain $\alpha_{p\pm}$. Incidentally, we can easily derive the relations of $g=\gamma=\beta_0$, $r_0=c/\beta_0$, $N=\kappa/\beta_0$, and $Nc=\kappa r_0$ from Case 3.
It is worth mentioning that Case 2 and Case 3 are not general case, because they require a large number of electrons (large $N$) or small enough distance $r_0$ to achieve strong coupling. Practically, Case 1 is a more general case when discussing coupled MNPs.

To make it clearer, we rewrite the solutions in Eq. (\ref{eq:alpha1}) in a simple form:
\begin{equation}
\begin{aligned}
\alpha_1=\alpha_{p-}=-\frac{\beta_1}{2}-\mathrm{i}\omega_1,~~
\alpha_2=\alpha_{p+}=-\frac{\beta_1}{2}+\mathrm{i}\omega_1,\\
\alpha_3=\alpha_{m-}=-\frac{\beta_2}{2}-\mathrm{i}\omega_2,~~
\alpha_4=\alpha_{m+}=-\frac{\beta_2}{2}+\mathrm{i}\omega_2.
\end{aligned}
\label{eq:alpha2}
\end{equation}
Here, there are two new eigenfrequencies ($\omega_1$ and $\omega_2$) and two new damping rates ($\beta_1$ and $\beta_2$). Obviously, they can be written as:
\begin{equation}
\begin{aligned}
\omega_1=\sqrt{\frac{\omega_0^2+g^2}{1+\eta}-\frac{\beta_1^2}{4}},~~
\beta_1=\frac{\beta_0+\gamma}{1+\eta},\\
\omega_2=\sqrt{\frac{\omega_0^2-g^2}{1-\eta}-\frac{\beta_2^2}{4}},~~
\beta_2=\frac{\beta_0-\gamma}{1-\eta}.
\end{aligned}
\label{eq:alpha3}
\end{equation}
The solutions of Eq. (\ref{eq:PLm}) for $y_j(t)$ can be written as:
\begin{equation}
y_j(t)=\sum_{k=1}^{4} B_{jk} \mathrm{exp}(\alpha_k t),~~
\mathrm{for}~j=1,2.
\label{eq:yt}
\end{equation}
There are 8 undetermined coefficients ($B_{jk}$), hence, we need 8 equations, i.e., 8 initial conditions. In Eq. (\ref{eq:PLm}), we list 4 of them, and the rest 4 conditions are determined by $\ddot{y}_j(0)$ and $\dddot{y}_j(0)$ which can be derived from Eq. (\ref{eq:basic2}):
\begin{equation}
\begin{aligned}
&A_1(\omega_{ex})=A_2(\omega_{ex})=A_0(\omega_{ex}),\\
&\ddot{y}_1(0)=\ddot{y}_2(0)=-\frac{\omega_0^2+g^2}{1+\eta}A_0=\ddot{y}_0,\\
&\dddot{y}_1(0)=\dddot{y}_2(0)=-\frac{\beta_0+\gamma}{1+\eta}\ddot{y}_0=\dddot{y}_0.
\end{aligned}
\label{eq:conditions}
\end{equation}.
Here, $A_0$ could be simplified from Eq. (\ref{eq:A}):
\begin{equation}
A_0(\omega_{ex})=\frac{C_0}{\Omega+G}=\frac{-C_0/(1+\eta)}{(\omega_{ex}+\mathrm{i} \beta_1 /2)^2-\omega_1^2}.
\label{eq:A0}
\end{equation}
The solutions of $B_{jk}$ should be given by the matrix equation:
\begin{equation}
\begin{aligned}
\begin{pmatrix}
  1      &  1       & 1        & 1  \\
\alpha_1   & \alpha_2   & \alpha_3   & \alpha_4   \\
\alpha_1^2 & \alpha_2^2 & \alpha_3^2 & \alpha_4^2 \\
\alpha_1^3 & \alpha_2^3 & \alpha_3^3 & \alpha_4^3 \\
\end{pmatrix}
&\left(
\begin{array}{c}
B_{j1} \\
B_{j2} \\
B_{j3} \\
B_{j4}
\end{array}\right)
=
\left(
\begin{array}{c}
A_0         \\
0           \\
\ddot{y}_0  \\
\dddot{y}_0
\end{array}\right)\\
&\mathrm{for} ~j=1,2.
\end{aligned}
\label{eq:Bjk}
\end{equation}
The solutions of $B_{jk}$ are listed below:
\begin{equation}
\begin{aligned}
&B_{1k}=B_{2k}=\frac{\dddot{y}_0-A_0\prod \limits_{n\ne k}^4\alpha_n-\ddot{y}_0\sum \limits_{n\ne k}^4\alpha_n}
{\prod \limits_{n\ne k}^4(\alpha_k-\alpha_n)}\\
&=\frac{\frac{(\omega_0^2+g^2)(\beta_0+\gamma)}{(1+\eta)^2}-\prod \limits_{n\ne k}^4\alpha_n+\frac{\omega_0^2+g^2}{1+\eta}\sum \limits_{n\ne k}^4\alpha_n}
{\prod \limits_{n\ne k}^4(\alpha_k-\alpha_n)} A_0=B_{0k}.
\end{aligned}
\label{eq:Bjk2}
\end{equation}.
Therefore, the solutions of Eq. (\ref{eq:basic2}) are:
\begin{equation}
y_1(t)=y_2(t)=\sum \limits_{k=1}^{4} B_{0k}\mathrm{exp}(\alpha_k t)
=y_0(t),
\label{eq:y0t}
\end{equation}.
After removing the insignificant coefficients, the far electric field is:
\begin{equation}
\begin{aligned}
&E_{far}(t)=N_1 \ddot{y}_1(t)+N_2 \ddot{y}_2(t)=2N \ddot{y}_0(t)=\sum \limits_{k=1}^{4} B_{0k}^{\prime}\mathrm{exp}(\alpha_k t),\\
&\mathrm{with}~ .
\end{aligned}
\label{eq:Et}
\end{equation}
where $B_{0k}^{\prime}=2N \alpha_k^2 B_{0k}$.
Hence, the emission spectrum can be evaluated by:
\begin{equation}
I_{PL}(\omega)=\mathrm{Re}\left< \int_0^{\infty} E_{far}^{*}(t)E_{far}(t+\tau)~\mathrm{exp}(\mathrm{i}\omega \tau)\mathrm{d}\tau \right>.
\label{eq:IPL0}
\end{equation}
Similarly, after removing the insignificant coefficients, the total PL spectrum can be written as:
\begin{equation}
\begin{aligned}
I_{PL}^{total}(\omega)
&= \frac{\beta_1 \left| B_{01}^{\prime} \right|^2}{\left( \omega-\omega_1\right)^2+(\frac{\beta_1}{2})^2}
+ \frac{\beta_1 \left| B_{02}^{\prime} \right|^2 }{\left( \omega+\omega_1\right)^2+(\frac{\beta_1}{2})^2}\\
&+ \frac{\beta_2 \left| B_{03}^{\prime} \right|^2}{\left( \omega-\omega_2\right)^2+(\frac{\beta_2}{2})^2}
+ \frac{\beta_2 \left| B_{04}^{\prime} \right|^2}{\left( \omega+\omega_2\right)^2+(\frac{\beta_2}{2})^2}
\end{aligned}
\label{eq:IPL1}
\end{equation}
After ignoring the second and the fourth term due to the fact that they are far away from the resonance frequency, the PL spectrum can be evaluated as:
\begin{equation}
I_{PL}(\omega)= \frac{\beta_1 \left| B_{01}^{\prime} \right|^2}{\left( \omega-\omega_1\right)^2+(\frac{\beta_1}{2})^2}
+ \frac{\beta_2 \left| B_{03}^{\prime} \right|^2}{\left( \omega-\omega_2\right)^2+(\frac{\beta_2}{2})^2}
\label{eq:IPL2}
\end{equation}
Notice that for the identical case, $B_{03}^{\prime}=B_{04}^{\prime}=0$ can be derived from Eq. (\ref{eq:Bjk2}). Hence, the identical PL spectrum can be evaluated as:
\begin{equation}
I_{PL}^{identical}(\omega)= \frac{\beta_1 \left| B_{01}^{\prime} \right|^2}{\left( \omega-\omega_1\right)^2+(\frac{\beta_1}{2})^2}
\label{eq:IPL3}
\end{equation}

\section{\label{sec:Results}Results and discussions}
After preparing these formulas employing this coupling model, the optical properties of the coupled system could be obtained easily.

\subsection{\label{sec:Coefficients} Coupling coefficients}
First, we investigate the behavior of the coupling coefficients, i.e., $g$, $\gamma$, and $\eta$, in the identical case.

\begin{figure}[tb]
\includegraphics[width=0.48\textwidth]{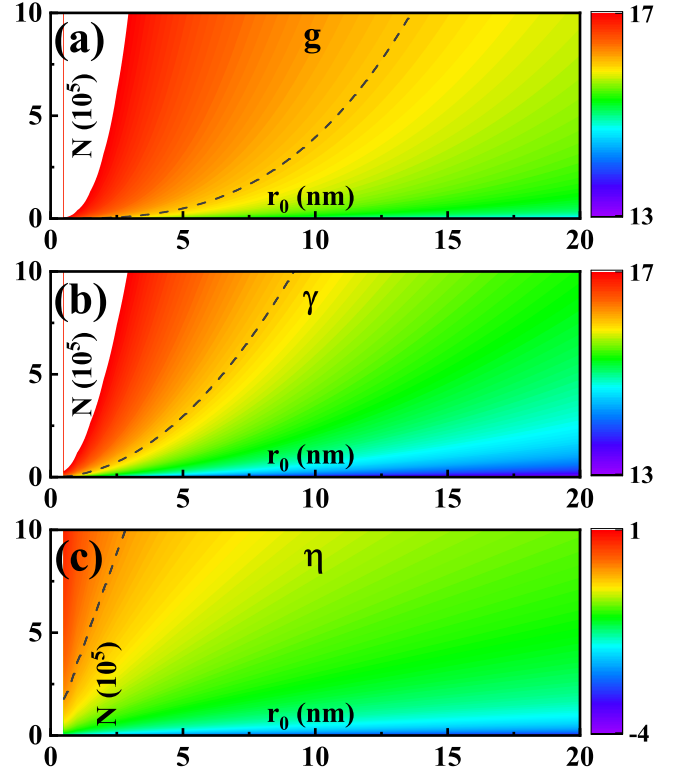}
\caption{\label{fig:CS} The coupling coefficients $g$ (a), $\gamma$ (b), and $\eta$ (c) varying with electron number $N$ and distance $r_0$. The color bar of (a) and (b) is in unit of Hz with log scale, i.e., $\mathrm{log}_{10}(g)$, and $\mathrm{log}_{10}(\gamma)$, respectively. Dashed lines stand for the contour line of the values $g,~\gamma=10^{16}$ Hz.
The color bar of (c) employs log scale, i.e., $\mathrm{log}_{10}(\eta)$. Dashed line stands for the contour line of the value $\eta=1$.
}
\end{figure}

Fig. \ref{fig:CS} shows these coupling coefficients varying with the electron number $N$ and the distance $r_0$, calculated from Eq. (\ref{eq:general}) and (\ref{eq:g}). As $N$ increases or/and $r_0$ decreases, they increase. The contour lines for $g$, $\gamma$, and $\eta$ satisfy $N \propto r_0^{1.5}$, $N \propto r_0^{2}$, and $N \propto r_0$, respectively. The dashed lines are the examples ($g=10^{16}$ Hz, $\gamma=10^{16}$ Hz, and $\eta=1$) to show the contour lines.

We emphasize here that when $r_0$ is small enough, i.e., $r_0<1$ nm, the electron tunneling effect makes it hard to evaluate the interaction between the oscillators employing only classical electromagnetic method which is used in this paper. Therefore, this model fails to describe the behaviors of the oscillators at the sub-nanometer scale, and we do not discuss this situation in this paper.

A particular case is when $\eta \ge 1$, which corresponds to a so strong coupling (very large $g$ and $\gamma$) that the practical MNPs cannot reach easily. Hence, this situation is not discussed in detail in this paper.

\subsection{\label{sec:Modes} Eigen modes}
Second, we investigate the eigen modes of the coupled system, i.e., $\omega_j$ and $\beta_j$ for $j=1,2$, in the identical case.

\begin{figure}[tb]
\includegraphics[width=0.48\textwidth]{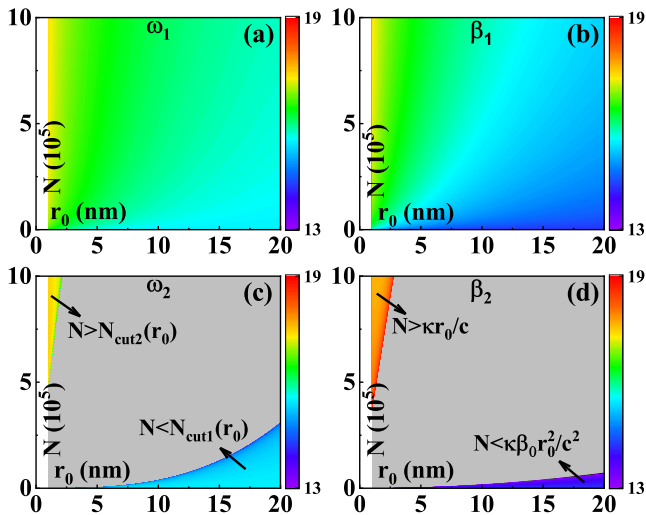}
\caption{\label{fig:OmegaBeta} Eigen modes of the coupled system. $\omega_1$ (a), $\beta_1$ (b), $\omega_2$ (c), and $\beta_2$ (d) varying with the electron number $N$ and distance $r_0$. Here, the parameters of the individual MNP are $\beta_0=1.57\times 10^{14}$ Hz (0.1034 eV), $\omega_0=3.14\times 10^{15}$ Hz (2.0690 eV). The gray region in (c) and (d) stand for the vanishing of the red mode.
}
\end{figure}

Fig. \ref{fig:OmegaBeta} shows the eigen modes varying with the electron number $N$ and the distance $r_0$, calculated from Eq. (\ref{eq:alpha3}). As the coupling coefficients increase, it is obvious that $\omega_1$ and $\beta_1$ (Mode 1) increase; however, $\omega_2$ and $\beta_2$ (Mode 2) decrease in the case of $\eta<1$, but they increase in the case of $\eta>1$. Due to the behaviours at $\eta<1$ (the general case), we could naturally call Mode 1 the blue branch or blue mode, and call Mode 2 the red branch or red mode.
Here, unit ``Hz'' and unit ``eV'' could be translated by $P(eV)=\frac{\hbar}{e}P(Hz)$, where $\hbar$ is the reduced Planck constant and $P$ is the parameter with unit ``Hz'' or ``eV''.

The blue mode could blue-shift along with the increasing damping coefficient as the coupling strength increases, which is only limited by the MNPs themselves, i.e., $N$ is limited by the volume and $r_0$ is limited by the geometric structures or the electron tunneling effect.
However, when it comes to the red mode, the behaviors are totally different. Both Fig. \ref{fig:OmegaBeta}c and \ref{fig:OmegaBeta}d show large areas filled with gray color, which indicates that the red mode vanishes in this area, i.e., only blue mode exists. The reason could be found in Eq. (\ref{eq:alpha3}).
Here, we take $\beta_2$ as an example to explain the reason. When $\eta<1$, i.e., the denominator of $\beta_2$ is positive, as the coupling strength increases from 0, the numerator of $\beta_2$ decreases from $\beta_0$ until $\gamma$ increases to $\beta_0$. After $\gamma$ continuous to increase, $\beta_2$ would be negative, indicating that the solutions $\alpha_{m\pm}$ should be abandoned due to its non-physical process (exponentially increase of $y_j(t)$). This region is indicated by $(\beta_0>\gamma) \cap (\eta<1)$, i.e., $N<\kappa \beta_0 r_0^2/c^2$, as shown at the bottom right corner in Fig. \ref{fig:OmegaBeta}d. When $\eta>1$, i.e., the denominator of $\beta_2$ is negative, hence, Mode 2 has physical meaning only when $\gamma$ is larger than $\beta_0$. As the coupling strength increases from a large value, $|\gamma-\beta_0|$ increases. This region is indicated by $(\beta_0<\gamma) \cap (\eta>1)$, i.e., $N>\kappa r_0/c$, as shown at the top left corner in Fig. \ref{fig:OmegaBeta}d.
Therefore, Mode 2 in the view of $\beta_2$ is forbidden at the gray region in Fig. \ref{fig:OmegaBeta}d, or Region $FR_{\beta2}:~ \kappa \beta_0 r_0^2/c^2< N < \kappa r_0/c$.
We analyze $\omega_2$ similarly. When the value of $\omega_2$ is an imaginary number, the value of $\alpha_{m\pm}$ would be a real number, indicating that Mode 2 is not an oscillation mode, thus abandoning Mode 2. Therefore, Mode 2 in the view of $\omega_2$ is forbidden at the gray region in Fig. \ref{fig:OmegaBeta}c, or Region $FR_{\omega2}:~ N_{cut1}< N < N_{cut2}$, where $N_{cut1} (r_0)$ and $N_{cut2}(r_0)$ are the two roots of the equation $(\omega_0^2-g^2)/(1-\eta)-\beta_2^2/4=0$, with $N$ the dependent variable and $r_0$ the independent variable.

According to the above analysis and the parameters ($\omega_0$ and $\beta_0$) we set in Fig. \ref{fig:OmegaBeta}, the total forbidden region of Mode 2 should be $FR_{\omega2} \cup FR_{\beta2}$, i.e., $\kappa \beta_0 r_0^2 / c^2 < N < N_{cut2} $.

\subsection{\label{sec:Spectra} Spectra}
Third, we investigate the spectra of the coupled system, i.e., white light scattering spectra, absorption spectra, and PL spectra. Here, we choose $g$ from $g,\gamma,\eta$ to represent the coupling strength for clarify.

\begin{figure}[tb]
\includegraphics[width=0.48\textwidth]{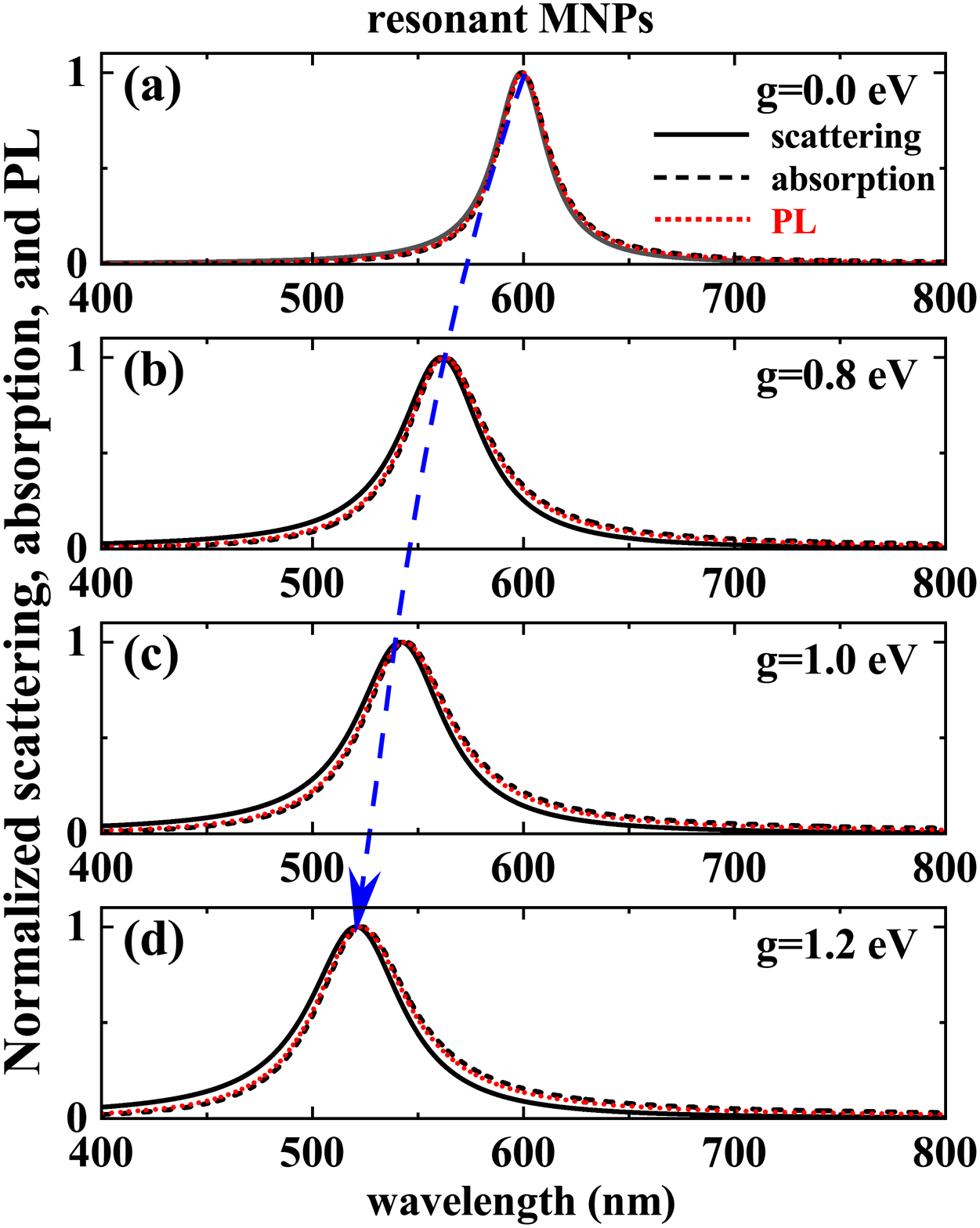}
\caption{\label{fig:Sca1} Normalized scattering (black solid), absorption (black dashed), and PL spectra (red dotted) for resonant MNPs, with different coupling strengths. (a)-(d) stand for $g=0.0$ eV, $g=0.8$ eV, $g=1.0$ eV, and $g=1.2$ eV, respectively. Blue dashed arrow stands for the track of the resonant peaks. Here, the parameters are $N_1=N_2=10^5$, $\beta_{01}=\beta_{02}=0.1034$ eV, $\omega_{01}=\omega_{02}=2.0690$ eV, and the excitation wavelength $\lambda_{ex}=355$ nm.
}
\end{figure}

Fig. \ref{fig:Sca1} shows these spectra for resonant MNPs varying with the coupling strength $g$, calculated from Eq. (\ref{eq:Isca1}), (\ref{eq:Iabs}), and (\ref{eq:IPL3}). Here, ``resonant MNPs'' stands for two identical MNPs, thus same resonant mode (both resonate at 600 nm). Two phenomena are illustrated. First, only one mode arises; second, the peaks blue shift and the line-width increases as $g$ increases.
The first phenomenon is due to the identical condition. In scattering and absorption view, Eq. (\ref{eq:A0}) shows no information of $\omega_2$, indicating the vanishing of Mode 2; in PL view, the amplitude of Mode 2 is zero according to Eq. (\ref{eq:IPL3}), also indicating the vanishing of Mode 2, thus only Mode 1 existing.
The second phenomenon could be explained by Eq. (\ref{eq:alpha3}) together with Fig. \ref{fig:OmegaBeta}a and \ref{fig:OmegaBeta}b, where $\omega_1$ and $\beta_1$ increase as $g$ increases. It is worth mentioning that this blue-shift phenomenon is opposite to the case of parallel polarization excitation as our previous work illustrates, where red-shift phenomenon is obtained.\cite{PLnjp} This is essentially caused by the fact that the coupling terms in the coupling equations [Eq. (\ref{eq:basic1})] of them have opposite sign, which are both derived from Eq. (\ref{eq:E}), i.e., positive sign for vertical polarization and negative sign for parallel polarization. Therefore, the primary peaks are different for these two cases according to Eq. (\ref{eq:A0}).
This phenomenon also agrees well with the experimental results of Pasquale et al.\cite{necklace}, where they employed several structures consisted with different numbers of MNPs including dimers. The experimental scattering spectra of the dimer show blue shift and red shift wit vertical and parallel polarization excitations, respectively.

\begin{figure}[tb]
\includegraphics[width=0.48\textwidth]{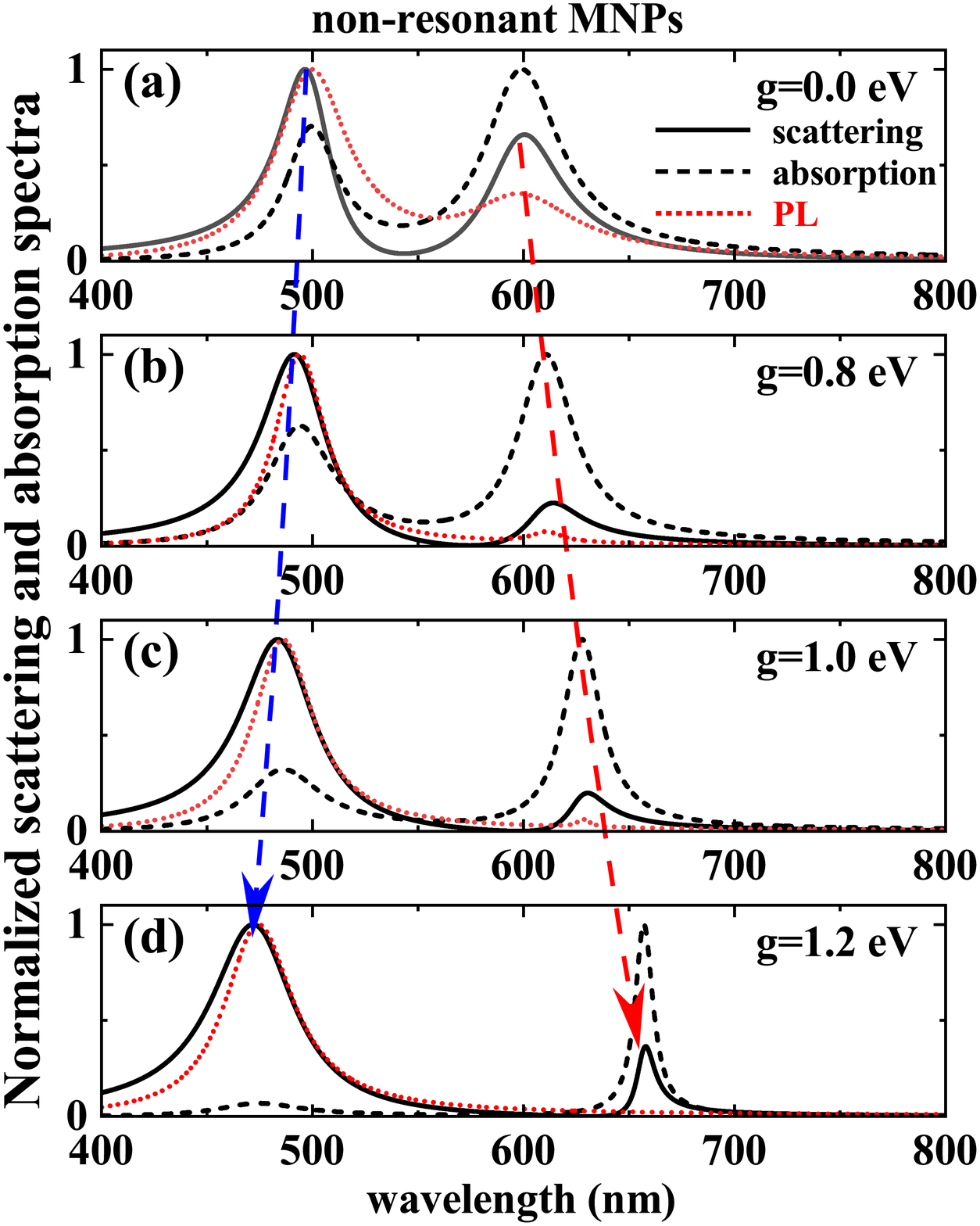}
\caption{\label{fig:Sca2} Normalized scattering (black solid), absorption (black dashed), and PL spectra (red dotted) for non-resonant MNPs, with different coupling strengths. (a)-(d) stand for $g=0.0$ eV, $g=0.8$ eV, $g=1.0$ eV, and $g=1.2$ eV, respectively. Blue and red dashed arrows stand for the tracks of the blue-shifted (Mode 1) and red-shifted (Mode 2) resonant peaks, respectively. $N_1=N_2=10^5$, $\beta_{01}=\beta_{02}=0.1034$ eV, $\omega_{01}=2.0718$ eV, $\omega_{02}=2.0690$ eV, and the excitation wavelength $\lambda_{ex}=355$ nm.
}
\end{figure}

Fig. \ref{fig:Sca2} shows these spectra for non-resonant MNPs varying with the coupling strength $g$, calculated from Eq. (\ref{eq:Isca1}) and (\ref{eq:Iabs}). Here, ``non-resonant MNPs'' stands for two different MNPs, thus different resonant modes (resonate at 500 nm and 600 nm, respectively). The formula to calculate the PL spectra is not shown in this paper because it is pretty complicated as has been mentioned. Also, two phenomena are illustrated. First, there are two resonant peaks with different ratios for these three; second, as $g$ increases, one peak blue shifts, the other red shifts. For the first phenomenon in detail, as $g$ increases, for the scattering and PL spectra, the ratio of the amplitude of Mode 1 to the amplitude of Mode 2 increases; however for the absorption spectra, the ratio decreases. Furthermore, the line-width of Mode 1 increases and the one of Mode 2 decreases with the increasing $g$, which could be explained by Eq. (\ref{eq:alpha3}) together with Fig. \ref{fig:OmegaBeta}b and \ref{fig:OmegaBeta}d. Similar to the ``resonant MNPs'' case, the second phenomenon could also be explained by Eq. (\ref{eq:alpha3}) and Fig. \ref{fig:OmegaBeta}a and \ref{fig:OmegaBeta}c.
In Fig. \ref{fig:Sca2}d, we notice that Mode 2 of PL vanishes. Because when $g=1.2$ eV, Mode 2 of PL reaches the forbidden region as discussed in Fig. \ref{fig:OmegaBeta}, thus only Mode 1 remaining.

\section{\label{sec:Conclusion}Conclusions}
In conclusion, we develop a coupling model based on classical electromagnetic method to explain the coupling properties of two MNPs with vertical polarization excitation. There are 3 coupling coefficients that influence the coupling and they vary with both the free electron number and the distance. The coupled scattering, absorption, and PL properties are illustrated. Particularly, the modes of PL are analyzed in detail. The identical MNPs case shows that the mode blue shifts as the coupling strength increases, which is opposite to the parallel polarization excitation (the mode red shifts as the coupling strength increases) that has been investigated in our previous work.\cite{PLnjp} This work would be helpful to understand the coupling properties of MNPs more deeply.

\section*{\label{sec:Acknowldegment}Acknowledgment}
This work was supported by the Fundamental Research Funds for the Central Universities (Grant No. FRF-TP-20-075A1).

\section*{Disclosures}
The authors declare no conflicts of interest.

\section*{Data availability}
The data that support the findings of this study are available from the corresponding author upon reasonable request.

\section*{\label{sec:Ref}References}

\bibliography{CM_Cheng}

\end{document}